\documentstyle[12pt]{article}

\newcommand{\be}{\begin{equation}}
\newcommand{\ee}{\end{equation}}
\newcommand{\bea}{\begin{array}}
\newcommand{\ea}{\end{array}}
\newcommand{\beqa}{\begin{eqnarray}}
\newcommand{\eeqa}{\end{eqnarray}}
\newcommand{\bean}{\begin{eqnarray*}}
\newcommand{\eean}{\end{eqnarray*}}
\newcommand{\eqn}[1]{(\ref{#1})}

\newcommand{\del}{\partial}

\newcommand{\C}{\:\mbox{\sf C} \hspace{-0.82em} \mbox{\sf C}\,}

\newcommand{\rom}[1]{\mbox{\rm #1}}
\def\Tr{\hbox{{\rom Tr}}}
\newcommand{\dd}[2]{\frac {\partial #1}{\partial #2}}
\newcommand{\nn}{\nonumber}

\def\up#1{\leavevmode \raise.16ex\hbox{#1}}
 
\def\sqr#1#2{{\vcenter{\vbox{\hrule height.#2pt
        \hbox{\vrule width.#2pt height#1pt \kern#1pt
          \vrule width.#2pt}
        \hrule height.#2pt}}}}

\newcommand{\journal}[4]{{\sl #1 }{\bf #2} \up(19#3\up) #4}

\newcommand{\gapproxeq}{\lower .7ex\hbox{$\;\stackrel{\textstyle >}{\sim}\;$}}
\newcommand{\lapproxeq}{\lower .7ex\hbox{$\;\stackrel{\textstyle <}{\sim}\;$}}
 
\newcounter{appendice}

\def\thebibliography#1{{\bf REFERENCES\markboth
 {REFERENCES}{REFERENCES}}\list
 {[\arabic{enumi}]}{\settowidth\labelwidth{[#1]}\leftmargin\labelwidth
 \advance\leftmargin\labelsep
 \usecounter{enumi}}
 \def\newblock{\hskip .11em plus .33em minus -.07em}
 \sloppy
 \sfcode`\.=1000\relax}

\begin{document}
\begin{flushright}
UAHEP-9603\\
DSFNA-T-9606\\ 
ESI-310\\
February 1996\\
\end{flushright}
\setcounter{footnote}{1}

\begin{center}
\vskip 0.5cm

{\large{\bf  Integrability of the Wess--Zumino--Witten  
model as a non--ultralocal theory}}

\vskip 0.8cm
$~^{*}${\bf S. G. Rajeev}\footnote{\scriptsize 
rajeev@urhep.pas.rochester.edu}, 
$~^{**}$ {\bf A. Stern}\footnote{\scriptsize astern@ua1vm.ua.edu} and 
$~^{***}${\bf P.~Vitale}\footnote{\scriptsize vitale@axpna1.na.infn.it}\\
{\it The Erwin Schr\"odinger Institute for Mathematical Physics\\
Pasteur gasse 6/7, A-1090 Wien Austria}\\
\vskip 0.4cm
$~^{*}${\it Department of Physics, University of Rochester\\
Rochester, NY 14627 USA}\\
\vskip 0.4cm
$~^{**}${\it Department of Physics, University of Alabama, 
Tuscaloosa, AL 35487 USA
}\\
\vskip 0.4cm
$~^{***}${\it Dipartimento di Scienze Fisiche, 
Universit\`a di Napoli\\ 
Mostra d'Oltremare, Pad.19 - 80125 
Napoli, ITALY}
\vskip 0.2cm
\end{center}
 
\vskip 0.8cm

\centerline{\bf Abstract}
 
We consider the 2--dimensional Wess--Zumino--Witten (WZW) model in the 
canonical 
formalism introduced in \cite{RSV}. Using an $r$--$s$ matrix approach to 
non--ultralocal field theories we find the Poisson algebra of monodromy 
matrices and of conserved quantities with a new, non--dynamical, $r$ 
matrix.
\vfill
\newpage

\section{Introduction}
 
In the last decades a large class of 2--dimensional solvable models in field 
theory has been well established both at the classical and at the quantum 
level, via the inverse scattering method; for a comprehensive review see 
\cite{Faddeev} and references therein. The distinguishing feature of such 
solvable models is the existence of a couple of matrices (a Lax pair), 
which are functions of the fields and of a spectral parameter, and which 
linearize
the equations of motion. Such matrices satisfy a Poisson algebra which is
ultralocal, which means that 
it doesn't contain derivatives of the delta function. 
It is this property which ensures the existence of a well defined Poisson 
bracket for monodromy matrices, 
the Jacobi identity for such an algebra being insured by 
the Yang--Baxter equation for an $r$--matrix. 

Ultralocality also implies that the $r$ matrix is non--dynamical (i.e., it 
doesn't depend on space--time coordinates). As a result an infinite subset 
of conserved quantities of the theory is in involution.

Unfortunately interesting theories like the WZW model are non--ultralocal. 
As was shown in \cite{Maillet}, non--ultralocality introduces discontinuous 
functions in 
the Poisson brackets of monodromy matrices, making it difficult to have a 
well defined algebra of conserved quantities; also, the Jacobi identity 
does not follow anymore from a Yang--Baxter equation. Nonetheless many 
attempts have been made to solve theories of this kind and solutions have 
been proposed for different models \cite{Maillet1}, \cite{Maillet}, 
\cite{Maillet3}, \cite{Abdalla1}, \cite{Abdalla2}, \cite{Abdalla3}, 
\cite{Konopelchenko}, \cite{Hlavaty}, \cite{Babelon}, 
\cite{Kundu},\cite{Semenov}.

In this paper we follow the approach 
contained in \cite{Maillet1} where an extended Yang--Baxter type algebra 
for the monodromy matrices is established and conditions for the conserved 
quantities to be in involution are given.

\section{Canonical formalism for the WZW model and associated linear 
system}

In \cite{RSV} a new canonical formalism for the WZW model was given. In 
this section we first sketch the conventional description, then 
summarize the results found in \cite{RSV} which will be used in the paper, 
and introduce the Lax pair for our theory.

The action which describes the model is 
\be
S =  \frac{1}{4 \chi^2}  \int d^2 x~~ 
\Tr (\del_\mu g  
\del_\nu g^{-1})\eta^{\mu\nu} \sqrt {-\rom{ det} (\eta)} + 
n \Gamma~~, \label{wzwact}
\ee 
where $g:R^{1,1} \longrightarrow SU(2)$ is a map from the two--dimensional
Minkowski space to the simple compact Lie group $SU(2)$,
$\eta$ is the Minkowskian metric, $\Gamma$ is the WZW action 
\be
     \Gamma= {1\over 24\pi}\int_B  d^3y \epsilon^{ijk}\Tr
g^{-1}\del_ig~g^{-1}\del_jg~g^{-1}\del_k g \label{Gamma}
\ee
and $B$ is a 3--dimensional manifold whose boundary is the space--time.
The coefficient of the WZW action is defined as 
\be
\rho=\frac{n \chi^2}{4\pi}.\label{ro} 
\ee
The equations of motion are 
\be
\del^\mu(\del_\mu g ~g^{-1}) -\rho \epsilon^{\mu\nu} \del_\mu g ~g^{-1}  
\del_\nu g~g^{-1}=0.  \label{eqmotion}
\ee 

The crucial property of this model is that it can be formulated entirely
in terms of the current $\del_{\mu}gg^{-1}$. The equation of motion
\eqn{eqmotion} is equivalent to the following pair of first order
equations, 
\beqa
     \dd{I}{t}&=&\dd{J}{x}+\rho[I,J]  \label{eqmo2}  \\
 \dd{J}{t}&=&\dd{I}{x}-[I,J]  \label{bianchi}
\eeqa
The second equation is the integrability condition for the existence of
$g:R^{1,1}\longrightarrow SU(2)$ satisfying, 
\beqa
  I &=&\dd{g}{t} g^{-1}  \label{defI}  \\
 J &=& \dd{g}{x} g^{-1} . \label{defJ}
\eeqa
$I$ and $J$ are traceless antihermitean matrices valued in the Lie algebra 
of $SU(2)$
\be
I=I_\alpha(x) {i\over 2} \sigma^\alpha,~~~~
J=J_\alpha (x){i\over 2} \sigma^\alpha
\label{IJ}
\ee
where $\sigma^\alpha$ are the Pauli matrices.
If we also impose the boundary condition
\be
     \lim_{x \rightarrow -\infty}g(x)=1  \label{bcg}
\ee
the solution  for $g$ is unique. Then equation \eqn{eqmo2} guarantees
that $g$ satisfy the equation of motion \eqn{eqmotion}. 

Though \eqn{eqmo2} can be
obtained from an action principle, the Hamiltonian formalism is more
suitable for our purposes; the Hamiltonian and Poisson brackets which
give \eqn{eqmo2} are then 
\be
     H_{1}=-\frac{1}{2\chi^{2}}\int tr(I^2 +J^2) dx~,
\label{ham1}
\ee
\beqa
 \frac{1}{2 \chi^{2}} \{ I_{\alpha}(x) ,I_{\beta}(y)
\}_{1}
&=&
     \varepsilon_{\alpha \beta \gamma} I_{\gamma}(x)\delta (x-y) +\rho
\varepsilon_{\alpha\beta\gamma}J_\gamma\delta(x-y)~,
\label{PBII}  \\
 \frac{1}{2 \chi^{2}}\{ I_{\alpha}(x) ,J_{\beta}(y)
\}_{1}
&=&
  \varepsilon_{\alpha \beta \gamma} J_{\gamma}(x)
     \delta (x-y) -\delta_{\alpha \beta} \delta^{\prime}(x-
y)~,
\label{PBIJ}\\
 \frac{1}{2 \chi^{2}}\{ J_{\alpha}(x) ,J_{\beta}(y)
\}_{1}
&=& 0~;\label{PBJJ}
\eeqa
where $\varepsilon_{\alpha\beta\gamma}$ are the structure constants of 
$SU(2)$.
$I_\alpha,~J_\alpha$ defined by \eqn{IJ} are square integrable functions 
on the space time;
the square integrability is a condition on how quickly the currents  must decay
to zero at infinity. It is needed for finiteness of energy (see
\eqn{ham1}).

As can be easily checked, the Poisson algebra given above is the
semidirect sum of an abelian algebra and a Kac--Moody algebra associated
to $SU(2)$. 

The same equations of motion can be obtained from an alternative 
Hamiltonian formalism \cite{RSV}, where the new Hamiltonian and Poisson 
brackets  depend on an extra  
parameter $\tau$, the limit $\tau \rightarrow 0$ being 
the conventional formalism. The alternative Hamiltonian and Poisson 
brackets are respectively:
\be
  H=-\frac{1}{2\chi^{2}(1-\tau^{2})^{2}}\int tr(I^2
+J^2) dx ~,  \label{ham2}
\ee
\beqa
 \frac{1}{2 \chi^{2}} \{ I_{\alpha}(x) ,I_{\beta}(y)\} &=&
     (1-\tau^2)\varepsilon_{\alpha \beta \gamma} I_{\gamma}(x)\delta (x-
y)\cr
& &+a(1-\tau^2)\varepsilon_{\alpha \beta \gamma}J_\gamma (x)\delta(x-y)
\label{PBII2}  \\
 \frac{1}{2 \chi^{2}}\{ I_{\alpha}(x) ,J_{\beta}(y)\} &=&
  (1-\tau^2)\varepsilon_{\alpha \beta \gamma} J_{\gamma}(x)
     \delta (x-y) \nn \\ 
& & -(1-\tau^2)^2\delta_{\alpha \beta}
\delta^{\prime}(x-y)\cr
& &+ (1-\tau^2)\epsilon \varepsilon_{\alpha \beta \gamma} I_\gamma
(x)\delta(x-y)
\label{PBIJ2}\\
 \frac{1}{2 \chi^{2}}\{ J_{\alpha}(x) ,J_{\beta}(y)\} &=&
     \tau^{2}(1-\tau^2)  \varepsilon_{\alpha \beta \gamma} 
I_{\gamma}(x)\delta(x-y)\cr
& &+(1-\tau^2)\mu \varepsilon_{\alpha\beta\gamma} J_\gamma(x)\delta(x-y)~,
\label{PBJJ2}
\eeqa
where $a, \mu, \epsilon$ are real parameters depending on $\tau$ as we 
will state below; $\tau$ can be chosen either 
real or imaginary as both the Hamiltonian and the Poisson brackets only 
depend on $\tau^2$. Let us call this algebra 
${\cal C}_2$.
It can be verified that we get the correct equations of motion, 
\eqn{eqmo2}, if we pose 
\be
     \rho={a-\epsilon\over 1-\tau^2}.
\ee
The Poisson algebra ${\cal C}_2$ can be rewritten in a much simpler 
way if we perform the change of variables: 
\beqa
     I&=&2\chi^{2}(1-\tau^{2})(\alpha L+\beta R)   \nn \\
     J&=&2 \chi^{2}(1-\tau^{2})(\gamma L+\delta R) .
\label{defR}
\eeqa
with $L$ and $R$ generators of two commuting Kac--Moody algebras, 
\beqa
\{ L_{\alpha}(x) ,L_{\beta}(y) \}_{2} &=&
  \varepsilon_{\alpha \beta \gamma} L_{\gamma}(x)
     \delta (x-y) +\frac{k}{2\pi}\delta_{\alpha \beta}
\delta^{\prime}(x-y)  \label{LL}\\
\{ R_{\alpha}(x) ,R_{\beta}(y) \}_{2} &=&
  \varepsilon_{\alpha \beta \gamma} R_{\gamma}(x)
     \delta (x-y) -\frac{\bar k}{2\pi}\delta_{\alpha \beta}
\delta^{\prime}(x-y)  \label{RR}\\
\{ L_{\alpha}(x) ,R_{\beta}(y) \}_{2} &=&0~,
\label{LR}
\eeqa
and $k,\bar k $ are a pair of constants.
 Note that the change of variables to $L$ and $R$ is singular if
$\tau=0$. 
It is now straightforward (although quite tedious) to show
that this algebra goes over to ${\cal C}_2$ under the above
change of variables, if we choose
\beqa
\alpha=1-\rho\tau\;\;& &
                \beta=1+\rho\tau\\
\gamma=\tau(\rho\tau-1)\;\;& &\delta=\tau(\rho\tau+1)\\
\epsilon=\mu=\rho\tau^2\;\;& & a=\rho \\
k={\pi\over 2\chi^2\tau(1-\rho\tau)^2}\;\; & &
\bar k={\pi\over 2\chi^2\tau(1+\rho\tau)^2}.
\eeqa
Thus our current algebra ${\cal C}_2$ is isomorphic to a direct
sum of two commuting Kac--Moody algebras, when $\tau$ is real.
We can conclude that the alternative canonical formalism for the WZW model 
here sketched is equivalent to the 
conventional one but has the advantage of exhibiting a Poisson algebra 
which is a sum of two Kac--Moody algebras,
$ \widehat{{\cal LSU}}(2) \oplus \widehat{{\cal LSU}}(2)$ 
if $\tau$ is real;
if we choose $\tau$ to be imaginary it can be checked that ${\cal C_2}$ 
is instead the Kac--Moody algebra associated to 
$SL(2, C)$, $ \widehat{{\cal LSL}}(2,C)$ (this is due to the fact that 
$I$ and $J$ become complex combinations of the 
$ \widehat{{\cal LSU}}(2) \oplus \widehat{{\cal LSU}}(2)$ generators, when 
$\tau$ is imaginary). We 
will assume from now on $\tau$ real, the procedure being identical for 
$\tau$ imaginary.

The WZW model admits an associated linear system \cite{Faddeev}
\beqa
\del_x \psi (x,t,\lambda)&=& A(x,t,\lambda) \psi (x,t,\lambda)\label{lin1}\\
\del_t \psi (x,t,\lambda)&=& M(x,t,\lambda) \psi (x,t,\lambda)~,\label{lin2}
\eeqa
where $(A,M)$ is the so called Lax pair, 
\beqa
A(x,t,\lambda)&=&
\frac{1}{2} I \biggl\{\frac{1-\rho}{1-\lambda} -
\frac{1+\rho}{1+\lambda}\biggr\} +
\frac{1}{2} J \biggl\{\frac{1-\rho}{1-\lambda} 
+\frac{1+\rho}{1+\lambda}\biggr\}\label{lx1}\\
M(x,t,\lambda)&=&
\frac{1}{2} I \biggl\{\frac{1-\rho}{1-\lambda} 
+\frac{1+\rho}{1+\lambda}\biggr\} +
\frac{1}{2} J \biggl\{\frac{1-\rho}{1-\lambda} 
-\frac{1+\rho}{1+\lambda}\biggr\}.
\label{lx2}
\eeqa
Because of their dependence on the currents, $A$ and  $M$ are 
traceless matrices valued in the Lie algebra of $SU(2)$ if the spectral 
parameter $\lambda$ is real, or in the Lie algebra of $SL(2, C)$ if we 
choose $\lambda$ to be complex. 
It is convenient to rewrite the Lax pair in terms of the new basis that 
we have chosen for the current algebra,
\be 
L(x,t)=\frac{\tau I -J}{4\chi^2\tau(1-\tau^2)(1-\rho\tau)}~,~~~~~~~~~~
R(x,t)=\frac{\tau I +J}{4\chi^2\tau(1-\tau^2)(1+\rho\tau)}~, 
\label{kacgen}
\ee
so that
\beqa
A(x,t,\lambda)&=& a(\lambda) L(x,t) + b(\lambda) 
R(x,t)\label{lax1}\\
M(x,t,\lambda)&=&c(\lambda) L(x,t) + d(\lambda) 
R(x,t)~,\label{lax2}
\eeqa
with
\beqa
a(\lambda)
&=& \chi^2(1-\tau^2) (1-\rho \tau) \biggl[ \frac{1-\rho}{1-\lambda} (1-\tau) - 
\frac{1+\rho}{1+\lambda} (1+\tau) \biggr] \nn\\
b(\lambda)
&=& \chi^2(1-\tau^2) (1+\rho \tau) \biggl[ \frac{1-\rho}{1-\lambda} (1+\tau) - 
\frac{1+\rho}{1+\lambda} (1-\tau) \biggr] \nn\\
c(\lambda)
&=& \chi^2(1-\tau^2) (1-\rho \tau) \biggl[ \frac{1-\rho}{1-\lambda} (1-\tau) + 
\frac{1+\rho}{1+\lambda} (1+\tau) \biggr] \nn\\
d(\lambda)
&=& \chi^2(1-\tau^2) (1+\rho \tau) \biggl[ \frac{1-\rho}{1-\lambda} (1+\tau) + 
\frac{1+\rho}{1+\lambda} (1-\tau) \biggr] .
\eeqa
The compatibility condition of equations \eqn{lin1} and \eqn{lin2} for 
any value of $\lambda$, which reads
\be
\del_t A-\del_x M +[A,M]=0 \label{0curv}
\ee
implies the equations of motion, \eqn{eqmo2} and 
\eqn{bianchi}. Relation \eqn{0curv} is also known as zero curvature 
condition for the connection $(M,A)$. 

We define the monodromy matrix $T(x,y,\lambda)$ (the t dependence being 
understood from now on), as a particular solution of \eqn{lin1}, 
\eqn{lin2}:
\beqa
\del_x T(x,y,\lambda)&=& A(x,t,\lambda) T (x,y,\lambda)\label{T1}\\
\del_t T(x,y,\lambda)&=& M(x,t,\lambda) T (x,y,\lambda)-
T (x,y,\lambda)M(y,t,\lambda)~, \label{T2}
\eeqa
with 
$$
T(x,x,\lambda)=1~~~~~~~T(y,x,\lambda)=T^{-1}(x,y,\lambda)
$$
$$
T(x,y,\lambda)T(y,z,\lambda)=T(x,z,\lambda).
$$
We have then
\be
T(x, y; \lambda)=
\rom{P} \exp \int_{y}^x A(x',\lambda) dx'
\ee
where P denotes path ordering.
The infinite volume limit of $T(x,y,\lambda)$
\be
T(\infty, -\infty; \lambda)\equiv T(\lambda)=
P \exp \int_{-\infty}^\infty A(x,\lambda) dx~,
\ee
is a conserved quantity for any value of $\lambda$. In fact from \eqn{T2} 
we have:
\be
\del_t T(\infty,-\infty;\lambda)=\del_t \biggl[M(\infty,\lambda)T(\lambda) - 
T(\lambda)M(-\infty,\lambda)\biggr] \label{conserv}
\ee
which is zero, because
$$
\lim_{x\rightarrow \pm \infty}I(x)=
\lim_{x\rightarrow \pm \infty}J(x)=0
$$
and $M$ a linear function of the currents $I,~J$. Note that for periodic 
boundary conditions only the trace of $T(\lambda)$ is conserved. 

From the definition of $T(\lambda)$ we note that $T(\lambda)$ (as a function 
of $\lambda$) is an element of the loop group of $SL(2, \C)$; in fact 
$A(\lambda=\pm \infty)=0$ thus implying
$$
T(\lambda=-\infty)=T(\lambda=\infty)=1.
$$
Also it satisfies
\be
 {T}^{\dag}(\lambda) T(\bar \lambda)=1.   \label{tdagt}
\ee
The last equation is easily proven observing that $A^{\dag}(x,\lambda)= 
-A(x, \bar\lambda)$
If we choose the spectral parameter to be real \eqn{tdagt} implies that 
$T(\lambda)$  be an element of the loop group of $SU(2)$.

Because $T(\lambda)$ is an ordered exponential of $A(x,\lambda)$, we only 
need the Poisson algebra of the $A's$ to determine the Poisson bracket of 
the monodromy matrices. 
From \eqn{lax1}, \eqn{lax2}, and the Poisson brackets of the Kac--Moody 
generators we get
\beqa
\{A^a(x,\lambda), A^b(y, \mu)\}&=&\varepsilon_{abc}\biggl(g(\lambda,\mu) 
A^c(x,\lambda) + g(\mu, \lambda)A^c(y, \mu)\biggr)\delta(x-y) \nonumber\\ 
 &-& f(\lambda,\mu)\delta_{ab} \delta'(x-y)~, \label{PoiA}
\eeqa
where 
\beqa
g(\lambda, \mu) &=& a(\mu) b(\mu) \frac{a(\lambda)- b(\lambda)}
{a(\lambda) b(\mu)-a(\mu) b(\lambda)} \\
g (\mu, \lambda) &=& -a(\lambda) b(\lambda) \frac{a(\mu)- b(\mu)}
{a(\lambda) b(\mu)-a(\mu) b(\lambda)}\\
f (\lambda,\mu) &=& b(\lambda) b(\mu) \frac{\bar k}{2\pi} - a(\lambda) 
a(\mu) \frac{k}{2\pi} .
\eeqa
We pose 
$$
\hat g( \lambda, \mu) = g( \lambda, \mu) \rom{C},~~~
\hat g( \mu, \lambda ) = g( \mu, \lambda ) \rom {C},~~~
\hat f( \lambda, \mu) = f( \lambda, \mu) \rom {C}
$$
where $\rom{C}={1\over 4}\sigma_a \otimes \sigma^a$.
Defining then $r(\lambda,\mu)$ and $s(\lambda,\mu)$ as the 
skew--symmetric and the symmetric part of $\hat g (\lambda, \mu)$, 
respectively,
\beqa
r(\lambda, \mu)&=&\frac{1}{2}\biggl(\hat g(\lambda, \mu)
-\hat g(\mu,\lambda)\biggr)\\
s (\mu, \lambda)&=&\frac{1}{2}\biggl(\hat g(\lambda, \mu)
+\hat g(\mu,\lambda)\biggr)~,
\eeqa
and observing that
\be
\hat f (\lambda,\mu)=2s (\lambda,\mu)
\ee
we can rewrite \eqn{PoiA} as
$$
\{A_1(x,\lambda), A_2(y, \mu)\}=\biggl([r(\lambda,\mu),  
A_1(x,\lambda) + A_2(x,\mu)] 
$$
\be
- [s(\lambda, \mu),A_1(x,\lambda) - 
A_2(x,\mu)]\biggr) \delta(x-y) 
- 2s(\lambda,\mu) \delta'(x-y) ~,\label{r-sA}
\ee
where $A= A^a {i\over 2}\sigma_a$, $A_1=A \otimes 1$, $A_2=1  \otimes A$. The 
matrices 
$r$, $s$, as explicit functions of $\lambda$ and $\mu$, are
\beqa
r&=& 2\chi^2 \frac{(1-\tau^2)}{\lambda-\mu} \biggl\{\frac{ 
\lambda^2(1-\rho^2\tau^2) -2\lambda\rho(1-\tau^2) 
+ \rho^2 -\tau^2} {\lambda^2-1} \nn\\
&+& \frac{ 
\mu^2(1-\rho^2\tau^2) -2\mu\rho(1-\tau^2) +\rho^2 -\tau^2} 
{\mu^2-1} \biggr\} \rom{C} \label{r}\\
s&=& 2\chi^2 \frac{(1-\tau^2)}{\lambda-\mu} \biggl\{\frac{ 
\lambda^2(1-\rho^2\tau^2) -2\lambda\rho(1-\tau^2) +\rho^2 -\tau^2} 
{\lambda^2-1} \nn\\
&-& \frac{ 
\mu^2(1-\rho^2\tau^2) -2\mu\rho(1-\tau^2) +\rho^2 -\tau^2} 
{\mu^2-1} \biggr\} \rom{C}~.\label{s}
\eeqa
In the limit $\rho=0, \tau=0$, which is the chiral model in the usual
formalism,\footnote{note that for $\rho=0, \tau\ne 0$ we have a family 
of principal chiral models depending on an extra parameter $\tau$ 
\cite{Rajeev}.} the $r$ and $s$ matrices given 
above reduce to the ones found in \cite{Maillet3} by general arguments on 
Poisson brackets related to Kac--Moody algebras.
Models which are described by a Poisson algebra containing derivatives of 
the delta function are called non--ultralocal \cite{Faddeev}. In general 
they also exhibit a space--time dependence for $r$ and $s$ matrices 
\cite{Maillet1}, \cite{Maillet},
the general form of the Poisson algebra thus being:
$$
  \{A_1(x,\lambda), A_2(y, \mu)\} = \biggl(\del_x r(x, \lambda,\mu) 
+[r(x,\lambda,\mu),  
A_1(x,\lambda) + A_2(x,\mu)] 
$$
\be
- [s(x,\lambda, \mu), A_1(x,\lambda) - 
A_2(x,\mu)]\biggr)\delta(x-y) 
-\biggl(s(x,\lambda,\mu)+s(y,\lambda,\mu)\biggr) \delta'(x-y)~.
\ee
In principle there could be higher derivatives of the delta function.
Note however that in our case, though the system is non--ultralocal, the 
$r$ and $s$ matrices are non--dynamical.
The Poisson bracket \eqn{r-sA} has a remarkable property: it is 
well defined for any value of $\tau$, even for the singular value $\tau=0$, 
where the Kac--Moody generators, L, R, 
are not independent functions of $I,~J$.

\section{Poisson algebra of monodromy matrices and 
equal-point limits}
In this section we summarize the procedure to obtain Poisson brackets for 
monodromy matrices starting from Poisson brackets of the currents 
$A(x,\lambda)$; we also find the equal--point limit and verify that our 
$r$ and $s$ matrices satisfy a sort of Yang--Baxter equation (see         
 \cite{Maillet}) which is the Jacobi identity for non--ultralocal theories. 
For the ultralocal case a detailed description is given 
in \cite{Faddeev}. 

The Poisson algebra for monodromy matrices is obtained from the 
algebra \eqn{r-sA} in the following way:
$$
\{T_{ab} (x,y,\lambda), T_{cd} (x',y',\mu)\} =
$$
$$
\int_y^x \int_{y'}^{x'} 
dz~dz' \frac{\delta T_{ab}(x,y,\lambda)} {\delta A_{ij}(z,\lambda)}
\{A_{ij}(z,\lambda), A_{kl}(z',\mu)\}        
\frac{\delta T_{cd}(x',y',\mu)}{\delta A_{kl}(z',\mu)} = 
$$
\be
 \int_y^x \int_{y'}^{x'} 
dz~dz' T_{ai}(x,z,\lambda) T_{jb}(z,y,\lambda) 
\{A_{ij}(z,\lambda), A_{kl}(z',\mu)\} 
       T_{ck}(x',z',\mu) T_{ld}(z',y',\mu)~, \label{coordform} 
\ee
where we used
\be
\delta T(x,y,\lambda)= \int_y^x T(x,z,\lambda) \delta A(z,\lambda) 
T(z,y,\lambda) dz.
\ee
Using tensorial formalism and the notation 
$T_1= T \otimes 1$, $T_2= 1 \otimes T$ 
we can rewrite \eqn{coordform} as
$$
\{T_1 (x,y,\lambda), T_2 (x',y',\mu)\} = 
$$
\be
\int_y^x \int_{y'}^{x'} 
dz~dz' T_1(x,z,\lambda) T_2 (x',z',\mu) 
\{A_1(z,\lambda), A_2(z',\mu)\} 
       T_1(z,y,\lambda) T_2(z',y',\mu).
\ee
Substituting \eqn{r-sA} and performing the integral we get
$$
\{T_1 (x,y,\lambda), T_2 (x',y',\mu)\} =
$$
\beqa
T_1(x,x_0,\lambda) T_2(x',x_0,\mu) \biggl(r (\lambda,\mu) &+& \epsilon 
(x-x') s(\lambda,\mu)\biggr) T_1(x_0,y,\lambda)T_2(x_0,y',\mu)-
\nn\\
T_1(x,y_0,\lambda) T_2(x',y_0,\mu) \biggl(r (\lambda,\mu) &+& \epsilon 
(y-y') s(\lambda,\mu)\biggr) T_1(y_0,y,\lambda)T_2(y_0,y',\mu)~,
\label{Tbra}
\eeqa
where $ \epsilon(x)=\rom{sign}(x)$, $x_0= \rom{min}(x, x')$, 
$y_0= \rom{max}(y, y')$.
We note that the Poisson brackets \eqn{Tbra} have been found under the 
explicit assumption that $x,~x',~y,~y'$, be all different; also this 
algebra shows up a discontinuity of amplitude $2s$, in the equal--point 
limits $x=x',~y=y'$.

The Jacobi identity for the Poisson brackets \eqn{Tbra} 
\be
\{T_1( x , y ,
\lambda), \{T_2( x', y',\mu), T_3( x'', y'', \nu) \}\} + \rom{cycl~ perm} =0  
\label{J}
\ee
results in an 
equation for the $r$ and $s$ matrices:
$$
[(r-s)_{12} (\lambda, \mu), (r+s)_{13}(\lambda, \nu)] +
$$
\be
[(r+s)_{12} (\lambda, \mu), (r+s)_{23}(\mu, \nu)] 
+ 
[(r+s)_{13} (\lambda, \nu), (r+s)_{23}(\mu, \nu)] =0~~.
  \label{YB}
\ee
If $r$ and $s$ depend on space--time variables, equation \eqn{YB} will 
also contain terms involving the Poisson bracket of $r$ and $s$ with $A$.
As can be seen \eqn{YB} reduces to the usual Yang--Baxter equation for
the $r$ matrix  when $s$ is zero. This is the ultralocal case. 

It can be checked that our $r$ and $s$ matrices given in \eqn{r} and 
\eqn{s} satisfy condition \eqn{YB},
which can also be rewritten as an equation for the matrix $g$.
Noting that
\be
\hat g(\lambda,\mu)=r(\lambda, \mu) + s(\lambda,\mu)~,~~~~
-\hat g(\mu, \lambda)=r(\lambda, \mu) - s(\lambda,\mu)
\ee
equation \eqn{YB} becomes then
$$
-[\hat g_{12} (\mu,\lambda), \hat g_{13}(\lambda, \nu)] +
$$
\be
+[\hat g_{12} (\lambda, \mu), \hat g_{23}(\mu, \nu)] + 
[\hat g_{13} (\lambda, \nu), \hat g_{23}(\mu, \nu)] =0~.
\ee
It can be verified, as a consistency check, that the same condition is 
obtained from the Jacobi identity of the current algebra \eqn{r-sA}.

\subsection{Equal-point limits of the monodromy algebra}
The monodromy  matrices $T(x,y,\lambda)$ evaluated at $x=\infty,~ 
y=-\infty$ are conserved quantities for any value of $\lambda$ as shown 
with \eqn{conserv}. We want to show that there are functions of 
them which are in involution with 
respect to the Poisson brackets \eqn{Tbra}. 
To do this, we have first to define the Poisson brackets 
\eqn{Tbra} for equal points $x=x',~ y=y'$. As already noted they are 
discontinuous at those points so that we cannot simply put $x=x',~ y=y'$. 
There are many regularization procedures to define such limit. We
follow here a  symmetric limit procedure illustrated in 
\cite{Maillet} which we briefly summarize in the following.
We pose 
\be
\Delta^{(1)}(x_1,x_2; y_1,y_2;\lambda_1,\lambda_2)=
\bigl\{T_1(x_1,y_1,\lambda_1), T_2(x_2,y_2,\lambda_2)\bigr\}
\ee
then the equal point limit $x_1=x_2=x$ is defined as:
\be
\Delta^{(1)}(x, x; y_1, y_2;\lambda_1, \lambda_2)=
\lim_{\epsilon\rightarrow 0} \frac{1}{2!} \sum_{\sigma} 
\Delta^{(1)}(x+\epsilon\sigma(1), x+\epsilon\sigma(2); y_1, y_2; \lambda_1, 
\lambda_2)~, \label{del1}
\ee
where the sum is over the permutations of $\{1, 2\}$. Analogously we can 
define the limit $y_1=y_2=y$, so that the Poisson bracket for two 
monodromy matrices at equal space--time points is 
\be
\{T_1 (x,y,\lambda), T_2 (x,y,\mu)\}= 
\Delta^{(1)}(x, x; y, y; \lambda, \mu). \label{EPB}
\ee
The introduction of the tensor $\Delta$ is particularly useful if 
generalized to n--nested Poisson brackets. In particular we are interested 
to rewrite the Jacobi identity using such a notation; we define
$$
\Delta^{(2)}(x_i,x_j, x_k; y_i,y_j, y_k;\lambda_i,\lambda_j, \lambda_k)=
\bigl\{T_i(x_i,y_i,\lambda_i), 
\bigl\{T_j(x_j,y_j,\lambda_j), T_k(x_k,y_k,\lambda_k)\bigr\}\bigr\}~,
$$
so that the symmetric limit $x_1=x_2=x_3$ is
$$
\Delta^{(2)}(x, x, x; y_i, y_j, y_k;\lambda_i, \lambda_j, \lambda_k)=
$$
\be
\lim_{\epsilon\rightarrow 0} \frac{1}{3!} \sum_{\sigma} 
\Delta^{(2)}(x+\epsilon\sigma(1), x+\epsilon\sigma(2), x+\epsilon\sigma(3);
y_1, y_2, y_3; \lambda_i, \lambda_j, \lambda_k). \label{del2} 
\ee
Using \eqn{EPB} to define the equal point limits for the monodromy matrices 
algebra we get
\be
\{T_1(x,y,\lambda), T_2(x,y,\mu)\}=[r(\lambda, \mu), T_1(x,y,\lambda) 
T_2(x,y,\mu)] \label{eqlim}.
\ee
Note that this Poisson bracket has the same form of the Poisson bracket 
obtained for ultralocal models but it satisfies Jacobi identity only 
through the symmetric limit procedure \eqn{del2}. 
Let us see this in more detail. 

Jacobi identity for \eqn{eqlim} is defined 
through the symmetric limit procedure \eqn{del2} as:
\be
\Delta^{(2)} (x,y;\lambda, \mu, \nu) +\Delta^{(2)} (x,y;\nu, \lambda, \mu
) + \Delta^{(2)} (x,y;\mu, \nu, \lambda ) =0 . \label{J2}
\ee 
This identity is implied from \eqn{J} (or 
equivalently \eqn{YB}).
Each term 
$$
\Delta^{(2)}(x,y;\lambda_i,\lambda_j,\lambda_k)=
$$
$$   
\lim_{\epsilon_1, \epsilon_2 \rightarrow 0} \sum_{\sigma, \tilde 
\sigma} \Delta^{(2)}\biggl( x +\epsilon_1 \sigma(1), x +\epsilon_1 \sigma(2), 
x +\epsilon_1 \sigma(3);y +\epsilon_2 \tilde\sigma(1), y+\epsilon_2 \tilde
\sigma(2), y+\epsilon_2 
\tilde\sigma(3);\lambda_i,\lambda_j,\lambda_k\biggr)
$$
is the sum of 36 terms, 6 for each choice of the x's configuration, so 
that we have 108 terms. They combine three by three in Jacobi identities 
of the kind \eqn{J}:
$$
\{T_1( x +\epsilon_1 \sigma(1), y +\epsilon_2 \tilde\sigma(1), 
\lambda_i), \{T_2( x +\epsilon_1 \sigma(2), y +\epsilon_2 \tilde\sigma(2), 
\lambda_j), T_3( x +\epsilon_1 \sigma(3), y +\epsilon_2 \tilde\sigma(3), 
\lambda_k) \}\} 
$$
$$
+ \rom{cycl~ perm} =0  
$$
because \eqn{J} is satisfied for each choice of $x\ne x' \ne x'',~ 
y\ne y' \ne y''$. Put 
$$
J_{\sigma(1)\sigma(2)\sigma(3)}= 
$$
\be
\sum_{\tilde
\sigma} \Delta^{(2)}(x +\epsilon_1 \sigma(1), x +\epsilon_1 \sigma(2), 
x +\epsilon_1 \sigma(3); y +\epsilon_2 \tilde\sigma(1), y+\epsilon_2 \tilde
\sigma(2), y+\epsilon_2 \tilde\sigma(3);\lambda_i,\lambda_j,\lambda_k)
\ee
(which we have just shown to be zero term by term); Jacobi identity for 
\eqn{eqlim} reads then
\be
\sum_{\sigma} J_{\sigma(1)\sigma(2)\sigma(3)}= 0.
\ee
If we check Jacobi 
identity directly on \eqn{eqlim} we find a Yang--Baxter equation for the 
$r$--matrix given in \eqn{r}, which is not satisfied. For this reason we 
say that \eqn{eqlim} satisfies Jacobi identity only in a weak sense.

In the infinite volume limit, equation \eqn{eqlim} reads
\be
\{T_1(\lambda), T_2(\mu)\}=[r(\lambda, \mu), T_1(\lambda)T_2(\mu)]~,
\label{tt}\ee
and the conserved quantities, ${\Tr}~ T(\lambda)$, 
are in involution, being ${\Tr}~ (A \otimes B)=  {\Tr}~ A \cdot
 {\Tr}~ B$, so that  
\be
\{{\Tr}~ T(\lambda), {\Tr}~ T(\mu)\}= \rom{Tr}~\{T_1(\lambda), 
T_2(\mu)\}~,
\ee
which is zero because of \eqn{tt}, being the trace of a commutator.
Note that the Poisson algebra
\be
\{{\Tr}~ T(\lambda), {\Tr}~ T(\mu)\}=0
\ee
satisfies Jacobi identity stronlgly (which is trivially true for zero
Poisson brackets).
 
\section{Conclusions}
We have found a three--parameter family of non--ultralocal 
 integrable models (the parameters being $\tau$, $\rho$, and 
the coupling constant $\chi$). The Poisson algebra of 
monodromy matrices can be rewritten in terms of $r$ and $s$ matrices 
which are independent on space--time variables and satisfy an extended 
Yang--Baxter equation. We also exhibit the conserved quantities of the 
theory which are in involution with respect to such a Poisson structure.
 
\section{Acknowledgements}
We wish to thank Giuseppe Marmo and Peter Michor for hospitality at ESI
where this work was done.
A.S. was supported in part
by the Department of Energy, USA, under contract number
DE-FG05-84ER40141.


\vfill
\newpage 

\begin{thebibliography}{99}
\bibitem{Faddeev} L. D. Faddeev, L.A. Takhtajan {\em Hamiltonian methods in the 
theory of solitons} Springer Verlag Ed. {\bf (1987)}
and refs. therein.
\bibitem{RSV} S. G. Rajeev, G. Sparano and P. Vitale \journal{Int J. of Mod. 
Phys.}{A31}{94}{5469} 
\bibitem{Rajeev} S. G. Rajeev \journal{Phys. Lett.}{B217}{89}{123}
\bibitem{Maillet1} J. M. Maillet \journal{Phys. Lett.}{B162}{85}{137}
\bibitem{Maillet} J. M. Maillet \journal{Nucl. Phys.}{B269}{86}{54}
\bibitem{Maillet3} J. M. Maillet \journal{Phys. Lett.}{B167}{86}{402}
\bibitem{Friedel} L. Friedel and J. M. Maillet \journal{Phys. Lett}{B262}
{91}{278}
\bibitem{Konopelchenko} B. G. Konopelchenko \journal{Rev. in Math. Phys}{2}
{90}{399}
\bibitem{Hlavaty} L. Hlavaty \journal{J. Math. Phys}{36}{95}{4883}
\bibitem{Babelon} O. Babelon and L. Bonora 
\journal{Phys. Lett}{B253}{91}{365}
\bibitem{Kundu} L. Hlavaty and A. Kundu {\sl Quantum integrability of 
non local models through Baxterization of quantized braided algebras}
 Preprint. hep-th 9406215.
\bibitem{Semenov} M. Semenov Tian-Shansky and A. Sevostyanov
{\sl Classical and quantum 
non--ultralocal systems on the lattice} Preprint. hep-th 9509029.
\bibitem{Abdalla1} M. C. B. Abdalla\journal{Phys. Lett.} {B152}{85}{215}
\bibitem{Abdalla2} E. Abdalla, M. C. B. Abdalla, J. C. Brunelli and
                  A. Zadra \journal{Comm. Math. Phys.} {166}{94}{379}
\bibitem{Abdalla3} E. Abdalla and M. Forger \journal{Mod. Phys. Lett.} 
                  {A7}{92}{2437} 
\end{thebibliography}
\end{document}